%% file: witten_index.tex
\documentclass[11pt]{article}
\usepackage{latexsym}
\usepackage{amssymb}
\usepackage{amsmath}
\usepackage{subfigure}
\usepackage[dvips]{graphicx}

\title{Analytic calculation of Witten index in $D=2$ supersymmetric Yang-Mills quantum mechanics}
\author{Piotr Korcyl\footnote{korcyl@th.if.uj.edu.pl}}
\date{M. Smoluchowski Institute of Physics, Jagiellonian University, Reymonta 4, 30-059 Krak\'{o}w, Poland}

\begin{document}

\maketitle

\begin{abstract}
We propose a method for the evaluation of Witten index in $D=2$ supersymmetric Yang-Mills quantum mechanics.
We rederive a known result for the $SU(2)$ gauge group and generalize it to any $SU(N)$ gauge group.
\end{abstract}

\section{Introduction}

        Witten index\cite{witten}, denoted by $I_W(T)$, was introduced as a tool to investigate the spontaneous supersymmetry
        breaking. The quantity
        \begin{equation}
        I_W(T) = \sum_i \big( e^{-T E^{\textrm{bosonic}}_i} - e^{-T E^{\textrm{fermionic}}_i} \big)
        \end{equation}
        has the advantage of being an example of an topological index. Hence, it may be calculated in perturbation theory
        in the weak coupling regime and continued to the strong coupling regime.
        As was argued in the original article by Witten\cite{witten}, $I_W(T)$ should be a nonzero integer in order for supersymmetry
        to be unbroken. The argument is straightforward for a system with discrete spectrum. All positive eigenenergies
        must be paired into supermultiplets and hence do not contribute to the Witten index. Contributions come only from the
        non-degenerate supersymmetric vacua. Hence, a nonzero value of $I_W(T)$ signifies the existence of such vacua and therefore
        points out that supersymmetry is not broken. In cases when $I_W(T)$ vanishes such argument is not conclusive. In this work
        we present an analytic evaluation of the Witten index of $D=2$ supersymmetric Yang-Mills quantum
        mechanics\cite{claudson}. Supersymmetric Yang-Mills quantum
        mechanics which are just dimensionally reduced a to single point in space quantum field theories
        have attracted a lot of attention due to their relations to, among other, the dynamics of supermembranes\cite{hoppe,dewit} or the
        dynamics of D0 branes in M-theory\cite{banks}.
        Up to now, the Witten index was calculated using very refined techniques
        for the higher dimensional models\cite{sethi+stern,staudacher}, namely with $D>2$, but not for $D=2$. A numerical
        Monte Carlo approach for evaluating $I_W(T)$ was recently discussed in Ref.\cite{kanamori}.


        In the following we start by briefly describing the $D=2$ supersymmetric Yang-Mills quantum
        mechanics and recalling the properties of their spectra. We concentrate on the degeneracy
        induced by the particle-hole symmetry and on the relation of the eigenenergies to the zeros of Laguerre
        polynomials. Consequently, in the following section we remind some analytic results concerning the asymptotic distribution of zeros
        of Laguerre polynomials, which turn out to be needed for the evaluation of $I_W(T)$.
        The calculation of $I_W(T)$ is presented in the subsequent sections.
        We analyze separately the models with $SU(2)$ and $SU(3)$ gauge symmetry and
        show some numerical results which support our approach.
        Then, we consider the generic models with any $SU(N)$ gauge symmetry and finish with some conclusions.

\section{Description of the systems}
\label{sec. opis}

        $D=2$ supersymmetric Yang-Mills quantum mechanics (SYMQM) were introduced by Claudson and Halpern\cite{claudson}.
        They may be thought of as systems obtained through dimensional reduction of quantum $D=2$ supersymmetric
        Yang-Mills quantum field theories to a single point in space\cite{doktorat_macka}. They consist
        of a set of $N^2-1$ real scalar fields and a set of $N^2-1$ fermion fields, both transforming in the
        adjoint representation of the $SU(N)$ group. They represent the simplest models of supersymmetric
        Yang-Mills quantum mechanics.

\subsection{Cut Fock basis approach}

        The cut Fock space approach to supersymmetric Yang-Mills quantum mechanics
        was proposed by Wosiek \cite{wosiek1} and described in details in Ref.\cite{korcyl1}. This approach was
        used to solve analytically the model with $SU(3)$ gauge group \cite{korcyl4,korcyl5} and subsequently the generic models with
        $SU(N)$ gauge symmetry \cite{korcyl6}.

        The Hilbert space of SYMQM models is composed of states invariant under
        global $SU(N)$ rotations as a consequence of the imposition of the dimensionally reduced Gauss law.
        It can be approximated by a subspace spanned by the Fock states having less than $N_{cut}$ quanta.
        We call $N_{cut}$ the cut-off. 
        For any finite value of $N_{cut}$ the spectra of all quantum systems are discrete. Hence, such cut-off can be
        used as a regularization of systems possessing continuous spectra. The physical eigenenergies
        are obtained in the $N_{cut} \rightarrow \infty$ limit.

        The basis states are constructed using the bosonic and fermionic bricks\cite{korcyl1}.
        For a given $N$ we define $N-1$ elementary bosonic bricks of the form (using the matrix notation),
        \begin{equation}
        C^{\dagger}_N(2) \equiv \textrm{tr}(a^{\dagger 2}), \ C^{\dagger}_N(3) \equiv \textrm{tr}(a^{\dagger 3}),
        \ \dots, \ C^{\dagger}_N(N) \equiv \textrm{tr}(a^{\dagger N}), \nonumber
        \end{equation}
        and any basis state is obtained from the Fock vacuum $|0\rangle$ as
        \begin{equation}
        |p_2, p_3, \dots, p_N\rangle =  C^{\dagger}_N(2)^{p_2} C^{\dagger}_N(3)^{p_3}  \dots  C^{\dagger}_N(N-1)^{p_{N-1}}  C^{\dagger}_N(N)^{p_N} |0\rangle.
        \label{eq. bozonowy stan bazowy sun}
        \end{equation}
        Indeed, the states eq.\eqref{eq. bozonowy stan bazowy sun} are linearly independent and provide
        a complete basis of the cut bosonic Hilbert space\cite{doktorat_macka}.

        In addition to the elementary bosonic bricks, one also has $d^{n_F}(N)$ composite fermionic bricks in the sector
        with $n_F$ fermionic quanta. We denote these bricks by $C^{\dagger}_N(n^{\alpha}_B, n_F, \alpha)$, where
        $n_B^{\alpha}$ corresponds to the number of bosonic creation operators and $n_F$ to the number of fermionic
        creation operators present in $C^{\dagger}_N(n^{\alpha}_B, n_F, \alpha)$. $\alpha$, $1 \le \alpha \le d^{n_F}(N)$,
        is an additional index needed to differentiate two fermionic bricks with equal numbers $n_B^{\alpha}$ and $n_F$.

        Fermionic basis states can be obtained through the action of appropriate fermionic bricks on the bosonic basis states
        eq.\eqref{eq. bozonowy stan bazowy sun},
        \begin{equation}
        |p_2, p_3, \dots, p_N, \alpha \rangle = C^{\dagger}_N(n^{\alpha}_B, n_F, \alpha) |p_2, p_3, \dots, p_N\rangle.
        \label{eq. fermionowy  stan bazowy sun}
        \end{equation}

\subsection{Spectra at finite cut-off}

        The Hamiltonian of SYMQM with $SU(N)$ symmetry reduces to a free Hamiltonian in the physical Hilbert space. It must be
        a $SU(N)$ singlet, so expressed in terms of creation and annihilation operators, it has the form,
        \begin{equation}
        H = (a^{\dagger}a) + \frac{N^2-1}{4} -\frac{1}{2} (a^{\dagger} a^{\dagger}) - \frac{1}{2}(aa).
        \label{eq. discrete hamiltonian}
        \end{equation}

        It was shown in Refs.\cite{korcyl5,korcyl6} that the spectra of these models for finite cut-off $N_{cut}$ are given exclusively by the zeros
        of appropriate generalized Laguerre polynomials. As it was described there, the set of all solutions can be
        divided into disjoint subsets called \emph{families}. These sets can be labeled by the maximal powers
        of elementary bricks present in the decomposition of energy eigenstates in the Fock basis. Thus, the solutions belonging
        to the family denoted by $(t_3, t_4, \dots, t_N)$ contain at most a $t_3$ power of the $(a^{\dagger 3})$ brick,
        a $t_4$ power of the $(a^{\dagger 4})$ brick and so on. The family $(0,0,\dots,0)$ contains solutions
        build of $(a^{\dagger 2})$ exclusively. In the fermionic sector the families must be labeled by
        an additional index $\alpha$, $(t_3, t_4, \dots, t_N; \alpha)$ which denotes the fermionic brick
        multipling the component proportional to the state $|t_2, t_3, \dots, t_N\rangle$.

        For a finite cut-off, each family contain a finite number of eigensolutions. The
        corresponding eigenenergies are given by quantization conditions, one per family.
        All quantization conditions have the same form, namely $L^{\gamma}_{m}(E)=0$, where
        $L^{\gamma}_m(x)$ is the generalized Laguerre polynomial of index $\gamma$,
        order $m$ and variable $x$.
        It was shown \cite{korcyl6} that the index $\gamma$ depends on the family labelling in the following way
        \begin{equation}
        \gamma = 3 t_3 + 4 t_4 + \dots + N t_N + n_B^{\alpha} + \frac{1}{2}(N^2-1) - 1,
        \label{eq. struktura indeksu}
        \end{equation}
        where the integers $n_B^{\alpha}$ denote the number
        of bosonic creation operators in the $\alpha$-th fermionic brick.

        The complete spectrum $\{E\}$ of the model with $SU(N)$ symmetry in the sector
        with $n_F$ fermionic quanta can be written in a compact form with the help
        of a polynomial $\Theta_{N_{cut}}^{n_F}(N,E)$, i.e.
        \begin{equation}
        \{E\} = \{E: \ \Theta_{N_{cut}}^{n_F}(N,E)= 0\},
        \end{equation}
        where \cite{korcyl6}
        \begin{multline}
        \Theta_{N_{cut}}^{n_F}(N,E) = \prod_{\alpha=1}^{d^{n_F}(N)} \Bigg\{
        \prod_{i=3}^N   \Bigg(  \\ \prod_{t_i=0}^{\big\lfloor \frac{1}{i} \big( N_{cut}- (\sum_{s=3}^{i} s t_s) - n_B^{\alpha}(N) \big) \big\rfloor}
        L_{\big\lfloor \frac{1}{2} \big( N_{cut}- (\sum_{s=3}^N s t_s) - n_B^{\alpha}(N) \big) \big\rfloor + 1}^{(\sum_{s=3}^N s t_s)  + \frac{1}{2}(N^2-1) -1 + n_B^{\alpha}(N)}(E) \Bigg) \Bigg\}.
        \label{eq. widmo fermionowe sun}
        \end{multline}
        The product over $\alpha$ corresponds
        to the contribution coming from every fermionic brick in the sector
        with $n_F$ fermionic quanta.
        $d^{n_F}(N)$ describes the number of independent fermionic
        bricks in sector with $n_F$ fermionic quanta and obviously must depend on $N$.
        Similarly, $n_B^{\alpha}(N)$ stands for the number of bosonic creation operators in the $\alpha$-th fermionic
        brick, which depends on both $n_F$ and $N$.
        The appearance of the integers $n_B^{\alpha}(N)$ is crucial for the analysis which follows.
        Next, there are $N-2$ products over the variables
        $t_3, \dots, t_N$. In fact, there are as many families
        as there are partitions of the numbers $0,1,2,\dots, N_{cut}-n_B^{\alpha}$ into the numbers $3,4,\dots,N$, which
        is taken into account by the upper limit of these products.

\section{Particle-hole symmetry}
\label{sec. particle-hole symmetry}

        Apart of supersymmetry, the SYMQM models have another symmetry which has important consequences for their spectra, namely
        the particle-hole symmetry. We describe the latter in this section.

        Hamiltonian eq.\eqref{eq. discrete hamiltonian} is a particular case of an operator invariant
        under the following transformation (see also Ref.\cite{shifman})
        \begin{equation}
        x \rightarrow - x, \quad p \rightarrow - p, \qquad f \rightarrow f^{\dagger}, \quad f^{\dagger} \rightarrow f.
        \label{eq. transformation p-h}
        \end{equation}
        By the same transformation the supercharges corresponding to eq.\eqref{eq. discrete hamiltonian}
        transform under \eqref{eq. transformation p-h} as
        $Q \rightarrow Q^{\dagger}$ and $Q^{\dagger} \rightarrow Q$.
        The canonical commutation and anticommutation relations remain unchanged,
        \begin{equation}
        [x_a, p_b] \rightarrow [-x_a, -p_b] = i \delta_{a,b}, \qquad \{ f^{}_a, f^{\dagger}_b \} \rightarrow \{ f^{\dagger}_a, f^{}_b \}= \delta_{a,b}.
        \end{equation}
        Therefore, there exist an unitary operator $U$ which realizes such transformation in the Hilbert space. Obviously,
        $U^2 = \mathcal{I}$, and hence, $U^{\dagger}=U$. We have,
        \begin{equation}
        U f_a^{\dagger} U = f^{}_a, \qquad U f^{}_a U = f_a^{\dagger}, \qquad U x_a U = - x_a, \qquad U p_a U = - p_a.
        \end{equation}

        One can check that the image of the Fock vacuum under $U$ is an eigenstate
        of the fermion occupation number operator $\textrm{tr} (f^{\dagger} f)$. Indeed,
        \begin{equation}
        \textrm{tr} (f^{\dagger} f) U |0\rangle = f^{\dagger}_a f^{}_a U |0\rangle = U f^{}_a U U f_a^{\dagger} U U |0\rangle = U f^{}_a f^{\dagger}_a |0\rangle = (N^2-1) U |0\rangle,
        \end{equation}
        where we used the fact that $a = 1, \dots, N^2-1$.
        Thus, the state $U |0\rangle$ is an eigenstate of $\textrm{tr} (f^{\dagger} f)$ to the
        eigenvalue $N^2-1$ which is the maximal value allowed by the Pauli exclusion principle.
        We denote such state by $|1\rangle \equiv U |0\rangle$.

        A generic bosonic state $|E\rangle_0$ can be written as
        \begin{equation}
        |E\rangle_0 = \sum_{n_B=0}^{\infty} f_{n_B}(a^{\dagger}; E) |0\rangle,
        \end{equation}
        with the coefficients $f_{n_B}(a^{\dagger}; E)$ being operators constructed with $n_B$-th power of the
        $a^{\dagger}$ operator, whereas numerical factors are chosen so that $H |E\rangle_0 = |E\rangle_0$.
        Therefore,
        \begin{equation}
        |E\rangle_{N^2-1} \equiv U|E\rangle_0 = \sum_{n_B=0}^{\infty} (-1)^{n_B} f_{n_B}(a^{\dagger}; E) |1\rangle.
        \label{eq. czynnik -1 w sektorze bozonowym}
        \end{equation}

        \begin{figure}[t]
        \begin{center}
        \subfigure[$SU(2)$]{
        \input{schemat_su2}
        \label{fig. struktura su2}
        }
        \subfigure[$SU(3)$]{
        \input{schemat_su3}
        \label{fig. struktura su3}
        }
        \end{center}
        \end{figure}
        \begin{figure}[!h]
        \begin{center}
        \subfigure[$SU(even)$]{
        \input{schemat_su4}
        \label{fig. struktura su_even}
        }
        \subfigure[$SU(odd)$]{
        \input{schemat_su5}
        \label{fig. struktura su_odd}
        }
        \caption{
        Schematic structure of supermultiplets in models with different $SU(N)$ gauge groups.
        Marks on the horizontal axis denote fermionic sectors with consecutive number of fermionic quanta.
        For the $SU(2)$ model (figure \ref{fig. struktura su2})
        $n_F=0,\dots,3$, whereas for the $SU(3)$ model (figure \ref{fig. struktura su3}) $n_F=0,\dots,8$. Figures \ref{fig. struktura su_even} and \ref{fig. struktura su_odd}
        show a generic situation for $N$ even and odd. Horizontal intervals connecting neighboring fermionic sectors
        represent possible supermultiplets constructed with states from these sectors. Note the degeneracy of the
        spectrum in the middle sector (with $n_F=\frac{1}{2}(N^2-1)$) in models with $N$ odd. The $SU(2)$ model is somewhat special, since
        parity forbids the connection of states coming from sectors with $n_F=1$ and $n_F=2$.\label{fig. struktura supermultipletow}}
        \end{center}
        \end{figure}
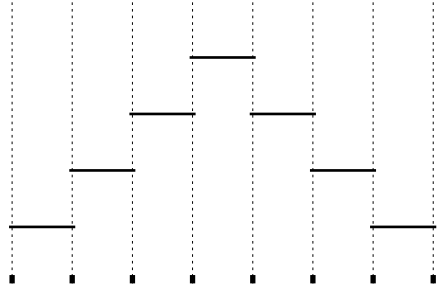
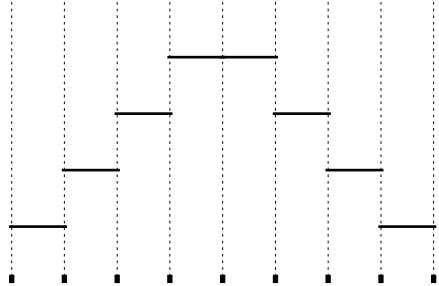

        A simple generalization of the above observation to the sector with $n_F$ fermionic quanta is the following.
        For a state $|E\rangle_{n_F}$ we have
        \begin{equation}
        |E\rangle_{n_F} = \sum_{n_B=0}^{\infty} f_{n_B,n_F}(a^{\dagger}, f^{\dagger}; E) |0\rangle,
        \end{equation}
        and the image of this state under $U$ has the following decomposition,
        \begin{equation}
        |E\rangle_{N^2-1-n_F} \equiv U|E\rangle_{n_F} = \sum_{n_B=0}^{\infty} (-1)^{n_B} f_{n_B,n_F}(a^{\dagger}, f; E) |1\rangle.
        \end{equation}
        If $|E\rangle_{n_F}$ is an energy eigenstate to the eigenvalue $E$, $H |E\rangle_{n_F} = E |E\rangle_{n_F}$,
        the energy of the state $|E\rangle_{N^2-1- n_F}$ is,
        \begin{equation}
        H |E\rangle_{N^2-1-n_F} = H U |E\rangle_{n_F} = U H |E\rangle_{n_F} = E U |E\rangle_{n_F} = E |E\rangle_{N^2-1 - n_F}.
        \end{equation}
        Hence, the particle-hole symmetry generates a double degeneracy of the spectrum. To each eigenenergy coming from the sector
        with $n_F$ fermionic quanta, $n_F \le \frac{1}{2}(N^2-1)$,
        corresponds an equal eigenenergy in the sector with $N^2-1-n_F$ fermionic quanta.

        It follows that for the models with $SU(N)$ gauge groups with $N$ odd the spectrum
        in the middle sector (with $n_F = \frac{1}{2}(N^2-1)$) has a double degeneracy.
        The states from this sector form supermultiplets with both the left neighboring sector
        (with $n_F=\frac{1}{2}(N^2-1)-1$) and the right neighboring sector (with $n_F=\frac{1}{2}(N^2-1)+1$).
        The particle-hole symmetry requires that the spectra of the latter two sectors were identical.
        Hence, the spectrum of the sector with $n_F = \frac{1}{2}(N^2-1)$ is doubly degenerate. There is no such
        effect for the models with $N$ even.
        A schematic figure \ref{fig. struktura supermultipletow} depicts the above arguments.
        The difference in the structure of supermultiplets in models with $N$ even and $N$ odd can
        be seen by comparing figures (\ref{fig. struktura su2} and \ref{fig. struktura su_even}) with
        (\ref{fig. struktura su3} and \ref{fig. struktura su_odd}).

        The above discussion has immediate consequences for the Witten index. 
        The double degeneracy of the
        spectrum implies that $I_W(T)$ vanishes for any $N$ even. In these cases the degenerate
        sectors with $n_F$ and $N^2-1-n_F$ fermions have opposite parities under
        $(-1)^{n_F}$. For $N$ odd, one can define the restricted Witten index, denoted by $I_W(T)_R$,
        which is the sum over a single copy of eigenenergies. Thus, $I_W(T)_R = \frac{1}{2} I_W(T)$
        for $N$ odd. Note that the case of $N=2$ is somehow special, since
        parity forbids the connection between sectors $n_F=1$ and $n_F=2$.
        Hence, $I_W(T)_R$ can be also defined for this model and indeed is nontrivial \cite{wosiek4}.
        The introduction of the cut-off does not break the
        particle-hole symmetry; therefore $I_W(T)_R$ is a well defined quantity also at any finite cut-off.

        We now proceed with the evaluation of $I_W(T)_R$ for any $N$ odd. However, before we present explicit computations
        we remind some properties of the distributions of zeros of Laguerre polynomials.

\section{Moments of the zeros distribution of Laguerre polynomials}

        The distribution of zeros of Laguerre polynomial of index $\gamma$ can be defined for a polynomial of any order $\mathcal{N}$ as\cite{dehesa}
        \begin{equation}
        \rho^{\gamma}_{\mathcal{N}}(E) = \frac{1}{\mathcal{N}} \sum_{i=1}^{\mathcal{N}} \delta\big(E - E_i(\mathcal{N})\big),
        \end{equation}
        where $L^{\gamma}_{\mathcal{N}}\big(E_i(\mathcal{N})\big)=0$.
        Let us also introduce the moments of $\rho^{\gamma}_{\mathcal{N}}(E)$ as
        \begin{equation}
        \mu^{\gamma}_n(\mathcal{N}) = \frac{1}{\mathcal{N}} \sum_{i=1}^{\mathcal{N}} \big( E_i(\mathcal{N})\big)^n = \int_{-\infty}^{\infty} E^n d \rho^{\gamma}_{\mathcal{N}}(E).
        \end{equation}
        Such moments can be computed recursively\cite{dehesa}
        \begin{align}
        \begin{split}
        \mu^{\gamma}_0(\mathcal{N}) &= 1 \\
        \mu^{\gamma}_1(\mathcal{N}) &= \mathcal{N} + \gamma \\
        \mu^{\gamma}_2(\mathcal{N}) &= \big(\mathcal{N} + \gamma \big) \big( 2 \mathcal{N} + \gamma -1 \big) \\
        &\vdots
        \end{split}
        \end{align}
        with the general term given by
        \begin{equation}
        \mu^{\gamma}_n(\mathcal{N}) = \big(2 \mathcal{N} + \gamma - n +1 \big) \mu^{\gamma}_{n-1}(\mathcal{N}) + \mathcal{N} \sum_ {t=1}^{n-2} \mu^{\gamma}_{n-1-t}(\mathcal{N}) \mu^{\gamma}_t(\mathcal{N})
        \label{eq. moments recursion}
        \end{equation}
        For large $\mathcal{N}$ the moments $\mu^{\gamma}_n(\mathcal{N})$
        can be approximated as
        \begin{equation}
        \mu^{\gamma}_n(\mathcal{N}) = x_n \mathcal{N}^n + \gamma y_n \mathcal{N}^{n-1} + \mathcal{O}(\mathcal{N}^{n-1}) + \gamma \mathcal{O}(\mathcal{N}^{n-2}),
        \end{equation}
        with $x_0=1$ and $y_0=0$, $y_1=1$.
        Inserting this into the recursion relation eq.\eqref{eq. moments recursion} we get two coupled recursions for $x_n$ and $y_n$,
        \begin{equation}
        x_n = 2 x_{n-1} + \sum_{t=1}^{n-2} x_{n-1-t} x_t,
        \label{eq. rec for xn}
        \end{equation}
        and
        \begin{equation}
        y_n = 2 y_{n-1} + x_{n-1} + 2\sum_{t=1}^{n-2} x_{n-1-t} y_t.
        \end{equation}
        Eq.\eqref{eq. rec for xn} can be rewritten using the fact that $x_0=1$ as
        \begin{equation}
        x_n = \sum_{t=0}^{n-1} x_{n-1-t} x_t,
        \end{equation}
        which is the recursion relation for the Catalan numbers,
        \begin{equation}
        x_n = \frac{1}{n+1}\binom{2n}{n},
        \end{equation}
        in agreement with an earlier result \cite{dehesa2}. The recursion relation for $y_n$ then becomes
        \begin{equation}
        y_n = \frac{1}{n}\binom{2n-2}{n-1} + 2\sum_{t=0}^{n-1} \frac{1}{n-t}\binom{2n-2t-2}{n-1-t} y_t.
        \label{eq. rec for yn}
        \end{equation}
        The solution for $y_n$ can be guessed to be
        \begin{align}
        y_n = \binom{2n-1}{n-1}.
        \end{align}
        To prove this, one can use the identity (5.62) from Ref.\cite{knuth}, which reads for $n$ integer and all real $r$,$s$ and $t$,
        \begin{equation}
        \sum_k \binom{tk+r}{k} \binom{tn-tk+s}{n-k} \frac{r}{tk+r} = \binom{tn+r+s}{n}.
        \end{equation}
        Hence, we have
        \begin{equation}
        \mu^{\gamma}_n(\mathcal{N}) = \frac{1}{n+1}\binom{2n}{n} \mathcal{N}^n + \gamma \binom{2n-1}{n-1} \mathcal{N}^{n-1} + \mathcal{O}(\mathcal{N}^{n-1}) + \gamma \mathcal{O}(\mathcal{N}^{n-2}).
        \label{eq. moments}
        \end{equation}
        We will use this result in the following sections.

\section{Evaluation of the Witten index}

        The evaluation of the Witten index of SYMQM systems is nontrivial since their spectra
        in all sectors are continuous. Therefore, one needs to introduce a regularization in order
        to define the Witten index in a mathematically correct way.
        In Ref.\cite{wosiek4} the model with $SU(2)$ symmetry was considered and the regularization
        was done by putting the system in a ball of radius $R$.
        At the end of the calculations the limit of $R \rightarrow \infty$ was taken.
        The regularization proposed in this note is motivated by the Fock space approach \cite{korcyl1}
        and it is done automatically by the cut-off.
        The regulator is the maximal number of quanta contained in the basis states, which we denoted by $N_{cut}$,
        and we should eventually take the limit of
        $N_{cut} \rightarrow \infty$.
        Note that, as was observed in \cite{maciek2}, such cut-off provides a infrared and ultraviolet regularization.
        We have
        \begin{multline}
        I_W(T) 
        = \lim_{N_{cut} \rightarrow \infty} \int_0^{\infty} e^{-E T} \Big( \delta^{\textrm{bosonic}}(N_{cut})\rho^{\textrm{bosonic}}_{N_{cut}}(E) + \\ - \delta^{\textrm{fermionic}}(N_{cut})\rho^{\textrm{fermionic}}_{N_{cut}}(E) \Big) dE,
        \end{multline}
        where $\rho^{\textrm{bosonic}}_{N_{cut}}(E)$ and $\rho^{\textrm{fermionic}}_{N_{cut}}(E)$ are the densities
        of bosonic and fermionic eigenenergies respectively at cut-off $N_{cut}$, whereas $\delta^{\textrm{bosonic}}(N_{cut})$ and
        $\delta^{\textrm{fermionic}}(N_{cut})$ denote the number of bosonic and fermionic states, respectively.

\subsection{Model with $SU(2)$ gauge symmetry}

        In this section we consider the model with $SU(2)$ gauge group. We present the analytic treatment and describe some numerical
        results supporting our approach.

\subsubsection{Analytic calculation}

        We start with the simplest situation, namely of the restricted Witten index in the $SU(2)$ model.
        Although $N=2$ is even, one can define $I_W(T)_R$ for this model (see section \ref{sec. particle-hole symmetry}).
        There are only two sectors that need
        to be considered, so
        \begin{equation}
        I_W^{N_{cut}}(T)_R = \sum_i e^{-E^{n_F=0}_i T} - e^{-E^{n_F=1}_i T}
        \end{equation}
        According to eq.\eqref{eq. widmo fermionowe sun} the eigenenergies in the bosonic
        sector are given by the zeros of $L_{\lfloor \frac{1}{2}N_{cut} \rfloor + 1}^{\frac{1}{2}}(E^{n_F=0})$,
        whereas the eigenenergies in the sector with one fermionic quantum are given by the zeros of
        $L_{\lfloor \frac{1}{2}(N_{cut}-1) \rfloor + 1}^{\frac{3}{2}}(E^{n_F=1})$.

        The idea of the approach is to work at given finite cut-off and
        expand the exponent into a Taylor series. For simplicity we assume $N_{cut}$ to be odd, $N_{cut} = 2M-1$, so that
        $\lfloor \frac{1}{2}N_{cut} \rfloor +1=M$ and $\lfloor \frac{1}{2}(N_{cut}-1) \rfloor +1= M$. We get
        \begin{align}
        I^{N_{cut}}_W(T)_R 
        &= \sum_{n=0}^{\infty} \frac{(-1)^n}{n!} T^n M \Bigg( \int_0^{\infty} E^n \rho^{\frac{1}{2}}_{M}(E) dE
        -  \int_0^{\infty} E^n  \rho^{\frac{3}{2}}_{M}(E) dE \Bigg),
        \label{eq. 28}
        \end{align}
        The integrals in eq.\eqref{eq. 28} are nothing but the moments of the zeros distribution of
        Laguerre polynomials with an appropriate $\gamma$ index, namely $\gamma=\frac{1}{2}$ for $n_F=0$ and
        $\gamma=\frac{3}{2}$ for $n_F=1$. Hence,
        \begin{equation}
        I^{N_{cut}}_W(T)_R = \sum_{n=1}^{\infty} \frac{(-1)^n}{n!} T^n  M
        \Big( \mu^{\frac{1}{2}}_n ( M )
        -  \mu^{\frac{3}{2}}_n ( M ) \Big),
        \end{equation}
        where the sum starts at $n=1$ since the zeroth term vanishes, $\mu^{\frac{1}{2}}_0(M) -  \mu^{\frac{3}{2}}_0(M)=1-1=0$.
        Inserting the expression for the moments eq.\eqref{eq. moments} all terms which are not proportional to $\gamma$ cancel.
        The leading non-vanishing term in $M$ is therefore proportional to the index of Laguerre polynomials. We get, for large $M$,
        \begin{multline}
        I^{N_{cut}}_W(T)_R = (\frac{1}{2}-\frac{3}{2}) \sum_{n=1}^{\infty} \frac{(-1)^n}{n!} T^n  M^{n} \binom{2n-1}{n-1} =\\= - \sum_{n=1}^{\infty} \frac{(-1)^n}{n!} T^n  M^{n} \binom{2n-1}{n-1}.
        \end{multline}
        The sum can be performed to yield
        \begin{equation}
        \sum_{n=0}^{\infty} \frac{(-1)^n}{n!} x^n \binom{2n-1}{n-1} = \frac{1}{2} e^{-2x} I_0(2x),
        \end{equation}
        where $I_0(x)$ is the modified Bessel function of the first kind.
        The zeroth term is nontrivial and can be evaluated as
        \begin{align}
        \lim_{n \rightarrow 0} \frac{1}{n!}\binom{2n-1}{n-1} = \lim_{n \rightarrow 0} \frac{\Gamma(2n)}{\Gamma(n) \Gamma(n+1)^2} =
        \lim_{n \rightarrow 0} \frac{2^{2n-\frac{1}{2}}}{\sqrt{2\pi}} \frac{\Gamma(n+\frac{1}{2})}{\Gamma(n+1)^2} = \frac{\Gamma(\frac{1}{2})}{2\sqrt{\pi}}  = \frac{1}{2}.
        \end{align}
        \begin{figure}[t]
        \begin{center}
        \input{witten_exact_su2_2_latex}
        \caption{Comparison at finite cut-off of $I^{N_{cut}}_W(T)$ evaluated by
        explicitly calculating all eigenenergies (solid line) and $I^{N_{cut}}_W(T)$ evaluated with approximated moments (dashed line)
        for four different cut-offs: $N_{cut} = 101$ for the lower pair of lines, $N_{cut} = 201,401$ and $N_{cut}=601$ for the upper pair of lines.
        \label{fig. numerical evidence su2 b}}
        \end{center}
        \end{figure}
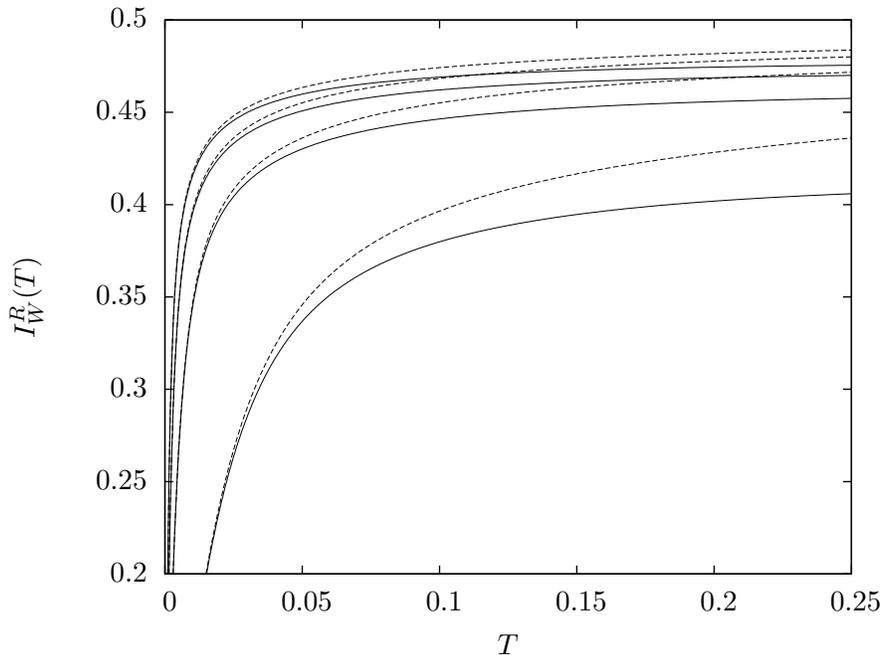
        Hence,
        \begin{equation}
        I^{N_{cut}}_W(T)_R = \frac{1}{2} - \frac{1}{2} e^{-(N_{cut}+1) T} I_0\big((N_{cut}+1) T\big),
        \label{eq. approximated witten index}
        \end{equation}
        One can exploit the asymptotic form of $I_0(x)$ for a large argument,
        \begin{equation}
        I_0(x) = \frac{e^x}{\sqrt{2 \pi x}} (1 + \mathcal{O}(\frac{1}{x})), \qquad \textrm{as } x \rightarrow \infty,
        \end{equation}
        to finally get
        \begin{equation}
        I^{N_{cut}}_W(T)_R = \frac{1}{2} - \frac{1}{2\sqrt{2 \pi} (N_{cut} T)^{\frac{1}{2}}} + \mathcal{O}(\frac{1}{(N_{cut} T)^{\frac{3}{2}}}) 
        \label{eq. wzor}
        \end{equation}
        Therefore, in the limit of infinite cut-off we obtain a $T$-independent value $\frac{1}{2}$, in agreement with the result of \cite{wosiek4},
        \begin{equation}
        I^{\infty}_W(T)_R = \frac{1}{2}.
        \label{eq. restricted su2}
        \end{equation}
        Eq.\eqref{eq. wzor} indicates also that the results converge to the limiting value in a rather slow way, namely as $\frac{1}{\sqrt{N_{cut}}}$.

        It also follows from the above discussion that for the restricted Witten index defined in the sectors $n_F=2$ and $n_F=3$ we get $I_W(T)_R = -\frac{1}{2}$.
        Therefore, summing the two contributions we get
        \begin{equation}
        I^{\infty}_W(T) = 0,
        \end{equation}
        which is in agreement with our expectation based on the particle-hole symmetry.

\subsubsection{Numerical evidence}


        On figure \ref{fig. numerical evidence su2 b} we plot $I_W(T)_R$ obtained
        for four different odd values of the cut-off, $N_{cut} = 101,201,401$ and $N_{cut}=601$.
        The solid lines represent $I_W(T)_R$ calculated by a numerical solution
        of the quantization conditions $L_{\lfloor \frac{1}{2}N_{cut} \rfloor + 1}^{\frac{1}{2}}(E^{n_F=0})=0$ and
        $L_{\lfloor \frac{1}{2}(N_{cut}-1) \rfloor + 1}^{\frac{3}{2}}(E^{n_F=1})=0$, taking the exponent and summing all contributions.
        The dashed lines correspond to expression eq.\eqref{eq. approximated witten index}.

        Three remarks can be made:
        \begin{itemize}
        \item both sets of curves to converge to the analytic prediction $\frac{1}{2}$,
        \item the curves present less and less dependence on $T$ as $N_{cut}$ increases,
        \item the discrepancy between the exact result (solid lines) and approximated one (dashed lines) is decreasing as we go to higher cut-offs.
        \end{itemize}

        The vanishing of $I^{N_{cut}}_W(T)_R$ at $T \rightarrow 0$ is related to the fact that $I^{N_{cut}}_W(0)_R$ is equal to
        the difference in the number
        of states in sectors with $n_F=0$ and $n_F=1$ which is 0 for $N_{cut}$ odd. For $N_{cut}$ even $I^{N_{cut}}_W(0)_R=1$, since there
        is one additional bosonic state.
        In the large $N_{cut}$ limit there should be no difference whether for $N_{cut}$ is odd and even, apart the $T=0$ point.

        The results shown on figure \ref{fig. numerical evidence su2 b} confirm that the approximations made
        in order to derive expression eq.\eqref{eq. approximated witten index} are correct.

%

\subsection{Model with $SU(3)$ gauge symmetry}

        In this section we present the generalization of the discussion  presented above to the case of the
        model with $SU(3)$ symmetry.

\subsubsection{Analytic calculation}

        Although the spectra in all fermionic sectors of the $SU(3)$ model are also given by the zeros
        of the generalized Laguerre polynomials\cite{korcyl5}, their structure turns out to be much more
        complicated than that of the $SU(2)$ model (see section \ref{sec. opis}).
        As was already mentioned, the eigensolutions can be grouped into disjoint sets called \emph{families}.
        To each family corresponds a single quantization condition.
        Hence, the spectrum in each sector is given by the zeros
        of a set of several Laguerre polynomials with different indices and of different orders.
        Therefore, we can write the Witten index as
        \begin{multline}
        I^{\infty}_W(T) 
        = \lim_{N_{cut} \rightarrow \infty} \Big( \sum_{\eta \in \textrm{ bosonic families}} \sum_i e^{-E^{\eta}_i(N_{cut}) T} + \\
        - \sum_{\eta' \in \textrm{ fermionic families}} \sum_i e^{-E^{\eta'}_i(N_{cut}) T} \Big)=  \\
        = \lim_{N_{cut} \rightarrow \infty} \int_0^{\infty} e^{-E T}
        \sum_{\substack{\eta \in \textrm{ bosonic families} \\ \eta' \in \textrm{ fermionic families}}}
        \Big( \delta^{\eta}(N_{cut}) \rho^{\eta}_{N_{cut}}(E)+\\
        - \delta^{\eta'}(N_{cut}) \rho^{\eta'}_{N_{cut}}(E) \Big) dE,
        \end{multline}
        where $\eta$,$\eta'$ are some multi-indices labeling the families present in this model, whereas
        $\delta^{\eta}(N_{cut})$ and $\delta^{\eta'}(N_{cut})$ are the numbers of solutions
        belonging to the family $\eta$ and $\eta'$, respectively, and are functions of $N_{cut}$.
        Their exact form will be presented below.

        For the $SU(3)$ model the families can be labeled by two integers: $(p, \beta)$ (see also eq.\eqref{eq. struktura indeksu}).
        We now consider the contribution to $I_W(T)$ coming from two families characterized by $(p,\beta)$ and $(q,\alpha)$,
        \begin{equation}
        I^{N_{cut}}_{q,\alpha; p,\beta}(T) = \int_0^{\infty} e^{-E T} \Big[ \delta^{(q,\alpha)}(N_{cut})\rho^{(q,\alpha)}_{N_{cut}}(E)
        -  \delta^{(p,\beta)}(N_{cut}) \rho^{(p,\beta)}_{N_{cut}}(E) \Big] dE.
        \end{equation}
        It was shown in Ref.\cite{korcyl5} that (see also eq.\eqref{eq. struktura indeksu} and eq.\eqref{eq. widmo fermionowe sun})
        \begin{align}
        \delta^{(q,\alpha)}(N_{cut}) &= \big\lfloor \frac{1}{2} \big( N_{cut}- 3 q) - n_B^{\alpha} \big) \big\rfloor + 1, \nonumber \\
        \rho^{(q,\alpha)}_{N_{cut}}(E) &= \rho^{3 q  + \frac{1}{2}(N^2-1) -1 + n_B^{\alpha}}_{\big\lfloor \frac{1}{2} \big( N_{cut}- 3 q) - n_B^{\alpha} \big) \big\rfloor + 1}(E). \nonumber
        \end{align}
        We approximate the factors $\delta^{(q,\alpha)}(N_{cut})$ and $\delta^{(p,\beta)}(N_{cut})$ as
        \begin{align}
        \big\lfloor \frac{1}{2} \big( N_{cut}- 3 q - n_B^{\alpha} \big) \big\rfloor + 1 \approx \frac{1}{2}N_{cut}, \qquad
        \big\lfloor \frac{1}{2} \big( N_{cut}- 3 p - n_B^{\beta} \big) \big\rfloor + 1 \approx \frac{1}{2}N_{cut},
        \end{align}
        which is justified in the leading order of large $N_{cut}$.
        Introducing the moments of the distributions we get, keeping only the leading
        terms in $N_{cut}$,
        \begin{multline}
        I^{N_{cut}}_{q,\alpha; p,\beta}(T) \approx 
        \sum_{n=1}^{\infty} \frac{(-1)^n}{n!}T^n \frac{1}{2}N_{cut}\Big[
           \frac{n}{n+1} \Big( \big\lfloor \frac{1}{2} \big(- 3 q - n_B^{\alpha} \big) \big\rfloor +\\- \big\lfloor \frac{1}{2} \big(- 3 p - n_B^{\beta} \big) \big\rfloor \Big) \binom{2n}{n}
        +  \Big( 3 q  + n_B^{\alpha} - 3 p  - n_B^{\beta}\Big) \binom{2n-1}{n-1} \Big] \Big(\frac{1}{2} N_{cut}\Big)^{n-1}.
        \end{multline}
        We perform the sums to get
        \begin{multline}
        I^{N_{cut}}_{q,\alpha; p,\beta}(T) = \frac{3 p  + n_B^{\beta} - 3 q  - n_B^{\alpha}}{2}\Big( 1 - e^{- N_{cut} T} I_0\big( N_{cut} T \big) \Big) \\ -
        \Big( \big\lfloor \frac{1}{2} \big(- 3 q - n_B^{\alpha} \big) \big\rfloor - \big\lfloor \frac{1}{2} \big(- 3 p - n_B^{\beta} \big) \big\rfloor\Big) e^{- N_{cut} T} I_1\big( N_{cut} T \big),
        \end{multline}
        which in the $N_{cut} \rightarrow \infty$ limit yields
        \begin{equation}
        I^{\infty}_{q,\alpha; p,\beta}(T) = \frac{3 p  + n_B^{\beta} - 3 q  - n_B^{\alpha}}{2}.
        \end{equation}

        In order to evaluate $I_W(T)$ one has to:
        \begin{enumerate}
        \item check that there are exactly as many bosonic families as there are fermionic ones,
        \item sum the contributions from all such pairs.
        \end{enumerate}

        The first point can be shown as follows. We will discuss it for a general $SU(N)$ SYMQM model.
        The number of singlet basis states with given number
        of bosonic and fermionic quanta, denoted by $D_{n_B, n_F}$, can be obtained, following \cite{doktorat_macka}, from the
        generating function
        \begin{equation}
        G(a,b) = \sum_{n_B=0}^{\infty} \sum_{n_F=0}^{\infty} D_{n_B, n_F} a^{n_B} (-b)^{n_F}.
        \end{equation}
        It can be shown\cite{doktorat_macka} that $G(a,b)$ has a convenient integral representation, namely
        \begin{equation}
        G(a,b) = \frac{1}{N!} \Big( \frac{1-b}{1-a} \Big)^{N-1} \int_0^{2 \pi} \prod_{i=1}^N \frac{d \alpha_i}{2 \pi}
        \prod_{i\ne j} \big(1-e^{i(\alpha_i-\alpha_j)}\big) \frac{1-b e^{i(\alpha_i-\alpha_j)}}{1-a e^{i(\alpha_i-\alpha_j)}}.
        \label{eq. full generating function}
        \end{equation}
        Moreover, it turns out the $G(a,b)$ can be also written as \cite{doktorat_macka}
        \begin{equation}
        G(a,b) = \Big( \prod_{k=2}^N \frac{1}{(1-a^k)} \Big) \sum_{n_F=0}^{N^2-1} (-b)^{n_F} c_{n_F}(a).
        \end{equation}
        Setting $b=1$ we obtain the difference of the sum over the bosonic sectors and the sum over the fermionic sectors,
        \begin{equation}
        \Big( \prod_{k=2}^N (1-a^k) \Big) G(a,b) \Big|_{b=1} = \sum_{n_F \textrm{ even}} c_{n_F}(a) - \sum_{n_F \textrm{ odd}} c_{n_F}(a).
        \label{eq. roznica}
        \end{equation}
        The polynomials $c_{n_F}(a) \equiv \sum_{n=0} \chi_{n}(n_F) a^n$ contain all the information about the numbers $n_B^{\alpha}$.
        Namely, there are $\chi_{n}(n_F)$ fermionic bricks with $n$ bosonic creation operators in the sector with $n_F$ fermionic quanta.
        From eq.\eqref{eq. full generating function} one immediately sees that the left-hand side of eq.\eqref{eq. roznica} vanishes,
        \begin{equation}
        \sum_{n=0} \Big( \sum_{n_F \textrm{ even}}  \chi_{n}(n_F) - \sum_{n_F \textrm{ odd}} \chi_{n}(n_F) \Big) a^n  = 0.
        \end{equation}
        Hence,
        \begin{equation}
        \sum_{n_F \textrm{ even}}  \chi_{n}(n_F) = \sum_{n_F \textrm{ odd}} \chi_{n}(n_F).
        \end{equation}
        This means that there is an equal number of fermionic bricks with $n$ bosonic creation operators in the sectors with $n_F$ even as there are such
        operators in the sectors with $n_F$ odd. It follows from this that the set of numbers $n_B^{\alpha}$ from the sectors
        with $n_F$ even is equal to the set from the sectors with $n_F$ odd,
        \begin{equation}
        \Big\{ n_B^{\alpha} \Big\}_{n_F \textrm{ even}} = \Big\{ n_B^{\alpha} \Big\}_{n_F \textrm{ odd}}
        \label{eq. rowne zbiory}
        \end{equation}
        This can be directly checked with explicit values for the $n_B^{\alpha}$ since for the $SU(3)$ model they are known explicitly,
        i.e. in Ref.\cite{korcyl1} all fermionic
        bricks for this model were presented. In table \ref{eq. indices} we just summarize the resulting values of $n_B^{\alpha}$ in different fermionic
        sectors.

        In order to show that there are as many bosonic families as fermionic ones we can now consider the following quantity, denoted by $W$,
        \begin{equation}
        W = \sum_{n_F \textrm{ even}}\sum_{\alpha=1}^{d^{n_F}}\sum_{q=0}^{\big\lfloor \frac{1}{3}(N_{cut}-n_B^{\alpha}) \big\rfloor} -
        \sum_{n_F \textrm{ odd}}\sum_{\beta=1}^{d^{n_F}} \sum_{p=0}^{\big\lfloor \frac{1}{3}(N_{cut}-n_B^{\beta}) \big\rfloor}
        \label{eq. rowna ilosc stanow}
        \end{equation}
        Since the sets of numbers $\big\{ n_B^{\alpha} \big\}_{n_F \textrm{ even}}$ and $\big\{ n_B^{\alpha} \big\}_{n_F \textrm{ odd}}$
        are equal, therefore $W=0$. Hence, at any finite cut-off, there is exactly as many bosonic families of solutions as there are fermionic ones.

        As far as the second point is concerned we have
        \begin{equation}
        I^{\infty}_W(T) = \sum_{n_F \textrm{ even}}\sum_{\alpha=1}^{d^{n_F}}\sum_{q=0}^{\big\lfloor \frac{1}{3}(N_{cut}-n_B^{\alpha})\big\rfloor}
        \frac{3 q  + n_B^{\alpha}}{2} -
        \sum_{n_F \textrm{ odd}}\sum_{\beta=1}^{d^{n_F}}\sum_{p=0}^{\big\lfloor \frac{1}{3}(N_{cut}-n_B^{\beta})\big\rfloor}
        \frac{3 p  + n_B^{\beta}}{2}.
        \end{equation}
        Again, it follows from eq.\eqref{eq. rowne zbiory} that
        \begin{equation}
        I^{\infty}_W(T) = 0,
        \label{eq. su3 result}
        \end{equation}
        and
        \begin{equation}
        I^{\infty}_W(T)_R = 0.
        \label{eq. su3 results r}
        \end{equation}

        \begin{table}
        \begin{equation}
        \begin{array}{|c||cc|c|c|c|c|}
        \hline
        n_F=0 & & n_B^1 = 0 & & & & \\
        \hline
        n_F=1 & & & n_B^1=1 & n_B^2 = 2 & & \\
        \hline
        n_F=2 & & & n_B^1=1 & n_B^2=2 & \ n_B^{3,4}=3 & \\
        \hline
        n_F=3 & & n_B^1=0 & n_B^2=1 & n_B^{3,4}=2 & n_B^{5,6,7}=3 & \ n_B^8=4 \\
        \hline
        n_F=4 & & & n_B^{1,2}=1 & n_B^{3,4,5,6}=2 & \ n_B^{7,8}=3 & n_B^{9,10}=4 \\
        \hline
        n_F=5 & & n_B^1=0 & n_B^2=1 & n_B^{3,4}=2 & n_B^{5,6,7}=3 & \ n_B^8=4 \\
        \hline
        n_F=6 & & & n_B^1=1 & n_B^2=2 & \ n_B^{3,4}=3 & \\
        \hline
        n_F=7 & & & n_B^1=1 & n_B^2=2 & & \\
        \hline
        n_F=8 & & n_B^1=0 & & & & \\
        \hline
        \end{array} \nonumber
        \end{equation}
        \caption{The values of $n_B^{\alpha}$ for the $SU(3)$ group.
        \label{eq. indices}}
        \end{table}

\subsubsection{Numerical evidence}

        The results given by eqs.\eqref{eq. su3 result} and \eqref{eq. su3 results r} were confirmed numerically.
        The spectra at given cut-off were calculated using the cut Fock basis approach and
        a recursive algorithm to speed up the calculations \cite{korcyl1}.
        We indeed obtained  $I^{N_{cut}}_W(T) = 0$ and $I^{N_{cut}}_W(T)_R = 0$ for any $N_{cut}$.

\section{Higher groups}

        The structure of solutions, namely the grouping into families having similar
        characteristics, for example the index of the Laguerre polynomials,
        which was observed for the model with $SU(3)$ symmetry \cite{korcyl5}, persists also for higher $N>3$\cite{korcyl6}.
        Hence, again, the spectrum consists of eigenenergies coming from the quantization conditions of all families, as is also
        indicated by eq.\eqref{eq. widmo fermionowe sun}.
        Thus, we can write the Witten index as
        \begin{multline}
        I^{\infty}_W(T) = \lim_{N_{cut} \rightarrow \infty} \int_0^{\infty} e^{-E T}
        \sum_{\substack{\eta \in \textrm{ bosonic families} \\ \eta' \in \textrm{ fermionic families}}}
        \Big( \delta^{\eta}(N_{cut}) \rho^{\eta}_{N_{cut}}(E) + \\
        - \delta^{\eta'}(N_{cut}) \rho^{\eta'}_{N_{cut}}(E) \Big) dE. 
        \label{eq. witten index sun}
        \end{multline}
        This time, the multi-index $\eta$ is even more complicated, as the generic families of
        the $SU(N)$ model are labeled by a set of integers $t_k$, $3\le k \le N$,
        and additionally by the index $\alpha$ denoting the fermionic brick, which multiplies the
        basis state with maximal number of bosonic quanta\cite{korcyl6}.

        The equality \eqref{eq. rowne zbiory} was derived for models with any gauge group $SU(N)$. Hence, one can now easily
        show that in the sum in eq.\eqref{eq. witten index sun} there are as many families of solutions
        in the sectors with $n_F$ even as there are families in the
        sectors with $n_F$ odd by generalizing the argument around eq.\eqref{eq. rowna ilosc stanow}. Therefore, we can sum
        the contributions to $I^{\infty}_W(T)$ coming from all pairs of families (taking account of eq.\eqref{eq. struktura indeksu}) and we get
        \begin{multline}
        I^{\infty}_W(T) = \frac{1}{2} \sum_{n_F \textrm{ even}} \sum_{\alpha=1}^{d^{n_F}} \sum_{t_3,\dots,t_N=1}^{\infty} \Big(  \big(\sum_{i=3}^{N} i t_i \big)+n_B^{\alpha}(n_F) \Big) \\ -
        \frac{1}{2} \sum_{n_F \textrm{ odd}}  \sum_{\alpha=1}^{d^{n_F}} \sum_{t_3,\dots,t_N=1}^{\infty} \Big( \big(\sum_{i=3}^{N} i t_i \big)+n_B^{\alpha}(n_F) \Big) ,
        \label{eq. prefactor2}
        \end{multline}
        from which, 
        using again eq.\eqref{eq. rowne zbiory}, we immediately get
%
%
        \begin{equation}
        I^{\infty}_W(T) = 0,
        \end{equation}
        and
        \begin{equation}
        I^{\infty}_W(T)_R = 0.
        \label{eq. restricted iw}
        \end{equation}

\section{Conclusions}

        In this paper we calculated the Witten index as well as the restricted Witten index for the
        $D=2$ supersymmetric Yang-Mills quantum mechanics with any $SU(N)$ gauge group. $I^{\infty}_W(T)$ vanishes for all, even and odd, $N$.
        We evaluated the restricted Witten index as well, which also vanishes for all $N$ odd, except of the $SU(2)$ model for which we recovered the known result $I_W(T)_R=\frac{1}{2}$.

        The vanishing of the Witten index does not imply that the supersymmetry is broken in this model. By studying the
        structure of the solutions \cite{korcyl6} one can persuade himself that there exist several nondegenerate
        states with zero energy in each studied models. They appear in nonadjacent sectors, therefore cannot be
        linked by the action of supercharges. The vanishing $I_W(T)$ signifies that there
        is as many supersymmetric vacua in sectors with $n_F$ even as there such states in sectors with $n_F$ odd,
        which could have been already suspected
        from the results of Ref.\cite{korcyl6} and Ref.\cite{maciek_lie_groups}. However, since the spectra
        of these models are continuous, the confirmation of this fact by the explicit evaluation
        of the Witten index is nontrivial.


\end{document}

%% file: schemat_su2.tex
\begingroup
  \makeatletter
  \providecommand\color[2][]{%
    \GenericError{(gnuplot) \space\space\space\@spaces}{%
      Package color not loaded in conjunction with
      terminal option `colourtext'%
    }{See the gnuplot documentation for explanation.%
    }{Either use 'blacktext' in gnuplot or load the package
      color.sty in LaTeX.}%
    \renewcommand\color[2][]{}%
  }%
  \providecommand\includegraphics[2][]{%
    \GenericError{(gnuplot) \space\space\space\@spaces}{%
      Package graphicx or graphics not loaded%
    }{See the gnuplot documentation for explanation.%
    }{The gnuplot epslatex terminal needs graphicx.sty or graphics.sty.}%
    \renewcommand\includegraphics[2][]{}%
  }%
  \providecommand\rotatebox[2]{#2}%
  \@ifundefined{ifGPcolor}{%
    \newif\ifGPcolor
    \GPcolorfalse
  }{}%
  \@ifundefined{ifGPblacktext}{%
    \newif\ifGPblacktext
    \GPblacktexttrue
  }{}%
  \let\gplgaddtomacro\g@addto@macro
  \gdef\gplbacktext{}%
  \gdef\gplfronttext{}%
  \makeatother
  \ifGPblacktext
    \def\colorrgb#1{}%
    \def\colorgray#1{}%
  \else
    \ifGPcolor
      \def\colorrgb#1{\color[rgb]{#1}}%
      \def\colorgray#1{\color[gray]{#1}}%
      \expandafter\def\csname LTw\endcsname{\color{white}}%
      \expandafter\def\csname LTb\endcsname{\color{black}}%
      \expandafter\def\csname LTa\endcsname{\color{black}}%
      \expandafter\def\csname LT0\endcsname{\color[rgb]{1,0,0}}%
      \expandafter\def\csname LT1\endcsname{\color[rgb]{0,1,0}}%
      \expandafter\def\csname LT2\endcsname{\color[rgb]{0,0,1}}%
      \expandafter\def\csname LT3\endcsname{\color[rgb]{1,0,1}}%
      \expandafter\def\csname LT4\endcsname{\color[rgb]{0,1,1}}%
      \expandafter\def\csname LT5\endcsname{\color[rgb]{1,1,0}}%
      \expandafter\def\csname LT6\endcsname{\color[rgb]{0,0,0}}%
      \expandafter\def\csname LT7\endcsname{\color[rgb]{1,0.3,0}}%
      \expandafter\def\csname LT8\endcsname{\color[rgb]{0.5,0.5,0.5}}%
    \else
      \def\colorrgb#1{\color{black}}%
      \def\colorgray#1{\color[gray]{#1}}%
      \expandafter\def\csname LTw\endcsname{\color{white}}%
      \expandafter\def\csname LTb\endcsname{\color{black}}%
      \expandafter\def\csname LTa\endcsname{\color{black}}%
      \expandafter\def\csname LT0\endcsname{\color{black}}%
      \expandafter\def\csname LT1\endcsname{\color{black}}%
      \expandafter\def\csname LT2\endcsname{\color{black}}%
      \expandafter\def\csname LT3\endcsname{\color{black}}%
      \expandafter\def\csname LT4\endcsname{\color{black}}%
      \expandafter\def\csname LT5\endcsname{\color{black}}%
      \expandafter\def\csname LT6\endcsname{\color{black}}%
      \expandafter\def\csname LT7\endcsname{\color{black}}%
      \expandafter\def\csname LT8\endcsname{\color{black}}%
    \fi
  \fi
  \setlength{\unitlength}{0.0500bp}%
  \begin{picture}(3968.00,2834.00)%
    \gplgaddtomacro\gplbacktext{%
      \csname LTb\endcsname%
      \put(426,220){\makebox(0,0){\strut{}0}}%
      \csname LTb\endcsname%
      \put(1465,220){\makebox(0,0){\strut{}1}}%
      \csname LTb\endcsname%
      \put(2503,220){\makebox(0,0){\strut{}2}}%
      \csname LTb\endcsname%
      \put(3542,220){\makebox(0,0){\strut{}3}}%
    }%
    \gplgaddtomacro\gplfronttext{%
    }%
    \gplbacktext
    \put(0,0){\includegraphics{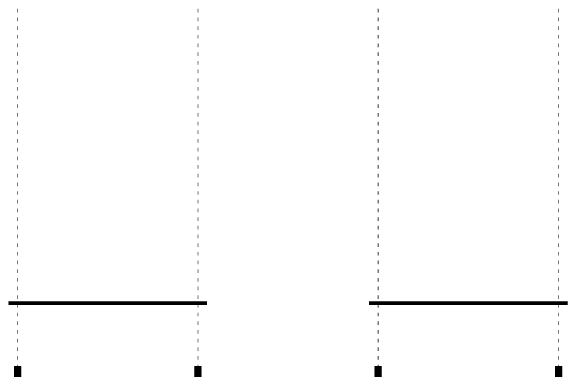}}%
    \gplfronttext
  \end{picture}%
\endgroup

%% file: schemat_su3.tex
\begingroup
  \makeatletter
  \providecommand\color[2][]{%
    \GenericError{(gnuplot) \space\space\space\@spaces}{%
      Package color not loaded in conjunction with
      terminal option `colourtext'%
    }{See the gnuplot documentation for explanation.%
    }{Either use 'blacktext' in gnuplot or load the package
      color.sty in LaTeX.}%
    \renewcommand\color[2][]{}%
  }%
  \providecommand\includegraphics[2][]{%
    \GenericError{(gnuplot) \space\space\space\@spaces}{%
      Package graphicx or graphics not loaded%
    }{See the gnuplot documentation for explanation.%
    }{The gnuplot epslatex terminal needs graphicx.sty or graphics.sty.}%
    \renewcommand\includegraphics[2][]{}%
  }%
  \providecommand\rotatebox[2]{#2}%
  \@ifundefined{ifGPcolor}{%
    \newif\ifGPcolor
    \GPcolorfalse
  }{}%
  \@ifundefined{ifGPblacktext}{%
    \newif\ifGPblacktext
    \GPblacktexttrue
  }{}%
  \let\gplgaddtomacro\g@addto@macro
  \gdef\gplbacktext{}%
  \gdef\gplfronttext{}%
  \makeatother
  \ifGPblacktext
    \def\colorrgb#1{}%
    \def\colorgray#1{}%
  \else
    \ifGPcolor
      \def\colorrgb#1{\color[rgb]{#1}}%
      \def\colorgray#1{\color[gray]{#1}}%
      \expandafter\def\csname LTw\endcsname{\color{white}}%
      \expandafter\def\csname LTb\endcsname{\color{black}}%
      \expandafter\def\csname LTa\endcsname{\color{black}}%
      \expandafter\def\csname LT0\endcsname{\color[rgb]{1,0,0}}%
      \expandafter\def\csname LT1\endcsname{\color[rgb]{0,1,0}}%
      \expandafter\def\csname LT2\endcsname{\color[rgb]{0,0,1}}%
      \expandafter\def\csname LT3\endcsname{\color[rgb]{1,0,1}}%
      \expandafter\def\csname LT4\endcsname{\color[rgb]{0,1,1}}%
      \expandafter\def\csname LT5\endcsname{\color[rgb]{1,1,0}}%
      \expandafter\def\csname LT6\endcsname{\color[rgb]{0,0,0}}%
      \expandafter\def\csname LT7\endcsname{\color[rgb]{1,0.3,0}}%
      \expandafter\def\csname LT8\endcsname{\color[rgb]{0.5,0.5,0.5}}%
    \else
      \def\colorrgb#1{\color{black}}%
      \def\colorgray#1{\color[gray]{#1}}%
      \expandafter\def\csname LTw\endcsname{\color{white}}%
      \expandafter\def\csname LTb\endcsname{\color{black}}%
      \expandafter\def\csname LTa\endcsname{\color{black}}%
      \expandafter\def\csname LT0\endcsname{\color{black}}%
      \expandafter\def\csname LT1\endcsname{\color{black}}%
      \expandafter\def\csname LT2\endcsname{\color{black}}%
      \expandafter\def\csname LT3\endcsname{\color{black}}%
      \expandafter\def\csname LT4\endcsname{\color{black}}%
      \expandafter\def\csname LT5\endcsname{\color{black}}%
      \expandafter\def\csname LT6\endcsname{\color{black}}%
      \expandafter\def\csname LT7\endcsname{\color{black}}%
      \expandafter\def\csname LT8\endcsname{\color{black}}%
    \fi
  \fi
  \setlength{\unitlength}{0.0500bp}%
  \begin{picture}(3968.00,2834.00)%
    \gplgaddtomacro\gplbacktext{%
      \csname LTb\endcsname%
      \put(394,220){\makebox(0,0){\strut{} 0}}%
      \csname LTb\endcsname%
      \put(791,220){\makebox(0,0){\strut{} 1}}%
      \csname LTb\endcsname%
      \put(1189,220){\makebox(0,0){\strut{} 2}}%
      \csname LTb\endcsname%
      \put(1586,220){\makebox(0,0){\strut{} 3}}%
      \csname LTb\endcsname%
      \put(1984,220){\makebox(0,0){\strut{} 4}}%
      \csname LTb\endcsname%
      \put(2382,220){\makebox(0,0){\strut{} 5}}%
      \csname LTb\endcsname%
      \put(2779,220){\makebox(0,0){\strut{} 6}}%
      \csname LTb\endcsname%
      \put(3177,220){\makebox(0,0){\strut{} 7}}%
      \csname LTb\endcsname%
      \put(3574,220){\makebox(0,0){\strut{} 8}}%
    }%
    \gplgaddtomacro\gplfronttext{%
    }%
    \gplbacktext
    \put(0,0){\includegraphics{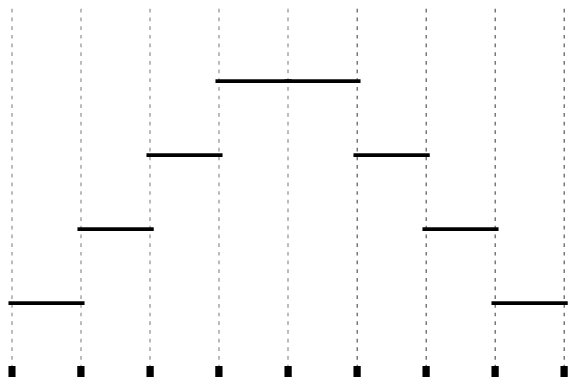}}%
    \gplfronttext
  \end{picture}%
\endgroup

%% file: schemat_su4.tex
\begingroup
  \makeatletter
  \providecommand\color[2][]{%
    \GenericError{(gnuplot) \space\space\space\@spaces}{%
      Package color not loaded in conjunction with
      terminal option `colourtext'%
    }{See the gnuplot documentation for explanation.%
    }{Either use 'blacktext' in gnuplot or load the package
      color.sty in LaTeX.}%
    \renewcommand\color[2][]{}%
  }%
  \providecommand\includegraphics[2][]{%
    \GenericError{(gnuplot) \space\space\space\@spaces}{%
      Package graphicx or graphics not loaded%
    }{See the gnuplot documentation for explanation.%
    }{The gnuplot epslatex terminal needs graphicx.sty or graphics.sty.}%
    \renewcommand\includegraphics[2][]{}%
  }%
  \providecommand\rotatebox[2]{#2}%
  \@ifundefined{ifGPcolor}{%
    \newif\ifGPcolor
    \GPcolorfalse
  }{}%
  \@ifundefined{ifGPblacktext}{%
    \newif\ifGPblacktext
    \GPblacktexttrue
  }{}%
  \let\gplgaddtomacro\g@addto@macro
  \gdef\gplbacktext{}%
  \gdef\gplfronttext{}%
  \makeatother
  \ifGPblacktext
    \def\colorrgb#1{}%
    \def\colorgray#1{}%
  \else
    \ifGPcolor
      \def\colorrgb#1{\color[rgb]{#1}}%
      \def\colorgray#1{\color[gray]{#1}}%
      \expandafter\def\csname LTw\endcsname{\color{white}}%
      \expandafter\def\csname LTb\endcsname{\color{black}}%
      \expandafter\def\csname LTa\endcsname{\color{black}}%
      \expandafter\def\csname LT0\endcsname{\color[rgb]{1,0,0}}%
      \expandafter\def\csname LT1\endcsname{\color[rgb]{0,1,0}}%
      \expandafter\def\csname LT2\endcsname{\color[rgb]{0,0,1}}%
      \expandafter\def\csname LT3\endcsname{\color[rgb]{1,0,1}}%
      \expandafter\def\csname LT4\endcsname{\color[rgb]{0,1,1}}%
      \expandafter\def\csname LT5\endcsname{\color[rgb]{1,1,0}}%
      \expandafter\def\csname LT6\endcsname{\color[rgb]{0,0,0}}%
      \expandafter\def\csname LT7\endcsname{\color[rgb]{1,0.3,0}}%
      \expandafter\def\csname LT8\endcsname{\color[rgb]{0.5,0.5,0.5}}%
    \else
      \def\colorrgb#1{\color{black}}%
      \def\colorgray#1{\color[gray]{#1}}%
      \expandafter\def\csname LTw\endcsname{\color{white}}%
      \expandafter\def\csname LTb\endcsname{\color{black}}%
      \expandafter\def\csname LTa\endcsname{\color{black}}%
      \expandafter\def\csname LT0\endcsname{\color{black}}%
      \expandafter\def\csname LT1\endcsname{\color{black}}%
      \expandafter\def\csname LT2\endcsname{\color{black}}%
      \expandafter\def\csname LT3\endcsname{\color{black}}%
      \expandafter\def\csname LT4\endcsname{\color{black}}%
      \expandafter\def\csname LT5\endcsname{\color{black}}%
      \expandafter\def\csname LT6\endcsname{\color{black}}%
      \expandafter\def\csname LT7\endcsname{\color{black}}%
      \expandafter\def\csname LT8\endcsname{\color{black}}%
    \fi
  \fi
  \setlength{\unitlength}{0.0500bp}%
  \begin{picture}(3968.00,2834.00)%
    \gplgaddtomacro\gplbacktext{%
      \csname LTb\endcsname%
      \put(397,220){\makebox(0,0){\strut{}}}%
      \csname LTb\endcsname%
      \put(850,220){\makebox(0,0){\strut{}}}%
      \csname LTb\endcsname%
      \put(1304,220){\makebox(0,0){\strut{}}}%
      \csname LTb\endcsname%
      \put(1757,220){\makebox(0,0){\strut{}}}%
      \csname LTb\endcsname%
      \put(2211,220){\makebox(0,0){\strut{}}}%
      \csname LTb\endcsname%
      \put(2664,220){\makebox(0,0){\strut{}}}%
      \csname LTb\endcsname%
      \put(3118,220){\makebox(0,0){\strut{}}}%
      \csname LTb\endcsname%
      \put(3571,220){\makebox(0,0){\strut{}}}%
    }%
    \gplgaddtomacro\gplfronttext{%
    }%
    \gplbacktext
    \put(0,0){\includegraphics{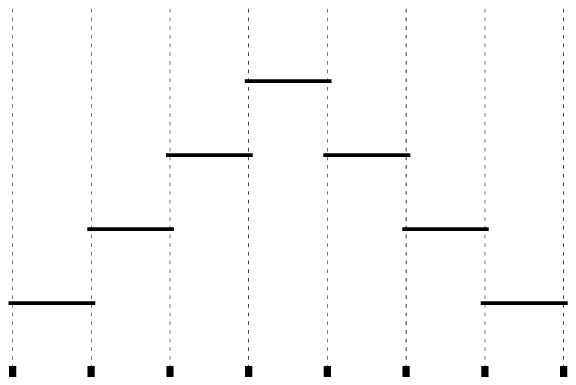}}%
    \gplfronttext
  \end{picture}%
\endgroup

%% file: schemat_su5.tex
\begingroup
  \makeatletter
  \providecommand\color[2][]{%
    \GenericError{(gnuplot) \space\space\space\@spaces}{%
      Package color not loaded in conjunction with
      terminal option `colourtext'%
    }{See the gnuplot documentation for explanation.%
    }{Either use 'blacktext' in gnuplot or load the package
      color.sty in LaTeX.}%
    \renewcommand\color[2][]{}%
  }%
  \providecommand\includegraphics[2][]{%
    \GenericError{(gnuplot) \space\space\space\@spaces}{%
      Package graphicx or graphics not loaded%
    }{See the gnuplot documentation for explanation.%
    }{The gnuplot epslatex terminal needs graphicx.sty or graphics.sty.}%
    \renewcommand\includegraphics[2][]{}%
  }%
  \providecommand\rotatebox[2]{#2}%
  \@ifundefined{ifGPcolor}{%
    \newif\ifGPcolor
    \GPcolorfalse
  }{}%
  \@ifundefined{ifGPblacktext}{%
    \newif\ifGPblacktext
    \GPblacktexttrue
  }{}%
  \let\gplgaddtomacro\g@addto@macro
  \gdef\gplbacktext{}%
  \gdef\gplfronttext{}%
  \makeatother
  \ifGPblacktext
    \def\colorrgb#1{}%
    \def\colorgray#1{}%
  \else
    \ifGPcolor
      \def\colorrgb#1{\color[rgb]{#1}}%
      \def\colorgray#1{\color[gray]{#1}}%
      \expandafter\def\csname LTw\endcsname{\color{white}}%
      \expandafter\def\csname LTb\endcsname{\color{black}}%
      \expandafter\def\csname LTa\endcsname{\color{black}}%
      \expandafter\def\csname LT0\endcsname{\color[rgb]{1,0,0}}%
      \expandafter\def\csname LT1\endcsname{\color[rgb]{0,1,0}}%
      \expandafter\def\csname LT2\endcsname{\color[rgb]{0,0,1}}%
      \expandafter\def\csname LT3\endcsname{\color[rgb]{1,0,1}}%
      \expandafter\def\csname LT4\endcsname{\color[rgb]{0,1,1}}%
      \expandafter\def\csname LT5\endcsname{\color[rgb]{1,1,0}}%
      \expandafter\def\csname LT6\endcsname{\color[rgb]{0,0,0}}%
      \expandafter\def\csname LT7\endcsname{\color[rgb]{1,0.3,0}}%
      \expandafter\def\csname LT8\endcsname{\color[rgb]{0.5,0.5,0.5}}%
    \else
      \def\colorrgb#1{\color{black}}%
      \def\colorgray#1{\color[gray]{#1}}%
      \expandafter\def\csname LTw\endcsname{\color{white}}%
      \expandafter\def\csname LTb\endcsname{\color{black}}%
      \expandafter\def\csname LTa\endcsname{\color{black}}%
      \expandafter\def\csname LT0\endcsname{\color{black}}%
      \expandafter\def\csname LT1\endcsname{\color{black}}%
      \expandafter\def\csname LT2\endcsname{\color{black}}%
      \expandafter\def\csname LT3\endcsname{\color{black}}%
      \expandafter\def\csname LT4\endcsname{\color{black}}%
      \expandafter\def\csname LT5\endcsname{\color{black}}%
      \expandafter\def\csname LT6\endcsname{\color{black}}%
      \expandafter\def\csname LT7\endcsname{\color{black}}%
      \expandafter\def\csname LT8\endcsname{\color{black}}%
    \fi
  \fi
  \setlength{\unitlength}{0.0500bp}%
  \begin{picture}(3968.00,2834.00)%
    \gplgaddtomacro\gplbacktext{%
      \csname LTb\endcsname%
      \put(394,220){\makebox(0,0){\strut{}}}%
      \csname LTb\endcsname%
      \put(791,220){\makebox(0,0){\strut{}}}%
      \csname LTb\endcsname%
      \put(1189,220){\makebox(0,0){\strut{}}}%
      \csname LTb\endcsname%
      \put(1586,220){\makebox(0,0){\strut{}}}%
      \csname LTb\endcsname%
      \put(1984,220){\makebox(0,0){\strut{}}}%
      \csname LTb\endcsname%
      \put(2382,220){\makebox(0,0){\strut{}}}%
      \csname LTb\endcsname%
      \put(2779,220){\makebox(0,0){\strut{}}}%
      \csname LTb\endcsname%
      \put(3177,220){\makebox(0,0){\strut{}}}%
      \csname LTb\endcsname%
      \put(3574,220){\makebox(0,0){\strut{}}}%
    }%
    \gplgaddtomacro\gplfronttext{%
    }%
    \gplbacktext
    \put(0,0){\includegraphics{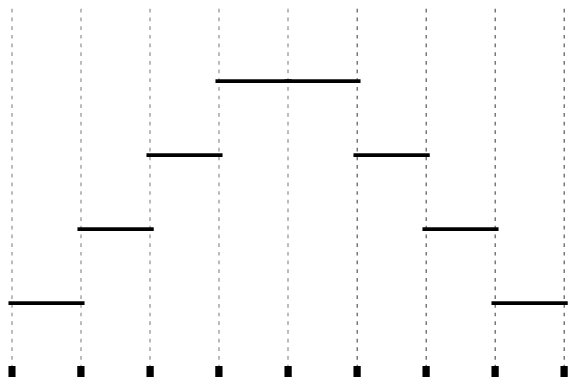}}%
    \gplfronttext
  \end{picture}%
\endgroup

%% file: witten_exact_su2_2_latex.tex
\begingroup
  \makeatletter
  \providecommand\color[2][]{%
    \GenericError{(gnuplot) \space\space\space\@spaces}{%
      Package color not loaded in conjunction with
      terminal option `colourtext'%
    }{See the gnuplot documentation for explanation.%
    }{Either use 'blacktext' in gnuplot or load the package
      color.sty in LaTeX.}%
    \renewcommand\color[2][]{}%
  }%
  \providecommand\includegraphics[2][]{%
    \GenericError{(gnuplot) \space\space\space\@spaces}{%
      Package graphicx or graphics not loaded%
    }{See the gnuplot documentation for explanation.%
    }{The gnuplot epslatex terminal needs graphicx.sty or graphics.sty.}%
    \renewcommand\includegraphics[2][]{}%
  }%
  \providecommand\rotatebox[2]{#2}%
  \@ifundefined{ifGPcolor}{%
    \newif\ifGPcolor
    \GPcolorfalse
  }{}%
  \@ifundefined{ifGPblacktext}{%
    \newif\ifGPblacktext
    \GPblacktexttrue
  }{}%
  \let\gplgaddtomacro\g@addto@macro
  \gdef\gplbacktext{}%
  \gdef\gplfronttext{}%
  \makeatother
  \ifGPblacktext
    \def\colorrgb#1{}%
    \def\colorgray#1{}%
  \else
    \ifGPcolor
      \def\colorrgb#1{\color[rgb]{#1}}%
      \def\colorgray#1{\color[gray]{#1}}%
      \expandafter\def\csname LTw\endcsname{\color{white}}%
      \expandafter\def\csname LTb\endcsname{\color{black}}%
      \expandafter\def\csname LTa\endcsname{\color{black}}%
      \expandafter\def\csname LT0\endcsname{\color[rgb]{1,0,0}}%
      \expandafter\def\csname LT1\endcsname{\color[rgb]{0,1,0}}%
      \expandafter\def\csname LT2\endcsname{\color[rgb]{0,0,1}}%
      \expandafter\def\csname LT3\endcsname{\color[rgb]{1,0,1}}%
      \expandafter\def\csname LT4\endcsname{\color[rgb]{0,1,1}}%
      \expandafter\def\csname LT5\endcsname{\color[rgb]{1,1,0}}%
      \expandafter\def\csname LT6\endcsname{\color[rgb]{0,0,0}}%
      \expandafter\def\csname LT7\endcsname{\color[rgb]{1,0.3,0}}%
      \expandafter\def\csname LT8\endcsname{\color[rgb]{0.5,0.5,0.5}}%
    \else
      \def\colorrgb#1{\color{black}}%
      \def\colorgray#1{\color[gray]{#1}}%
      \expandafter\def\csname LTw\endcsname{\color{white}}%
      \expandafter\def\csname LTb\endcsname{\color{black}}%
      \expandafter\def\csname LTa\endcsname{\color{black}}%
      \expandafter\def\csname LT0\endcsname{\color{black}}%
      \expandafter\def\csname LT1\endcsname{\color{black}}%
      \expandafter\def\csname LT2\endcsname{\color{black}}%
      \expandafter\def\csname LT3\endcsname{\color{black}}%
      \expandafter\def\csname LT4\endcsname{\color{black}}%
      \expandafter\def\csname LT5\endcsname{\color{black}}%
      \expandafter\def\csname LT6\endcsname{\color{black}}%
      \expandafter\def\csname LT7\endcsname{\color{black}}%
      \expandafter\def\csname LT8\endcsname{\color{black}}%
    \fi
  \fi
  \setlength{\unitlength}{0.0500bp}%
  \begin{picture}(6802.00,5102.00)%
    \gplgaddtomacro\gplbacktext{%
      \csname LTb\endcsname%
      \put(1122,660){\makebox(0,0)[r]{\strut{} 0.2}}%
      \put(1122,1356){\makebox(0,0)[r]{\strut{} 0.25}}%
      \put(1122,2053){\makebox(0,0)[r]{\strut{} 0.3}}%
      \put(1122,2749){\makebox(0,0)[r]{\strut{} 0.35}}%
      \put(1122,3445){\makebox(0,0)[r]{\strut{} 0.4}}%
      \put(1122,4142){\makebox(0,0)[r]{\strut{} 0.45}}%
      \put(1122,4838){\makebox(0,0)[r]{\strut{} 0.5}}%
      \put(1254,440){\makebox(0,0){\strut{} 0}}%
      \put(2289,440){\makebox(0,0){\strut{} 0.05}}%
      \put(3324,440){\makebox(0,0){\strut{} 0.1}}%
      \put(4358,440){\makebox(0,0){\strut{} 0.15}}%
      \put(5393,440){\makebox(0,0){\strut{} 0.2}}%
      \put(6428,440){\makebox(0,0){\strut{} 0.25}}%
      \put(220,2749){\rotatebox{90}{\makebox(0,0){\strut{}$I_W^R(T)$}}}%
      \put(3841,110){\makebox(0,0){\strut{}$T$}}%
    }%
    \gplgaddtomacro\gplfronttext{%
    }%
    \gplbacktext
    \put(0,0){\includegraphics{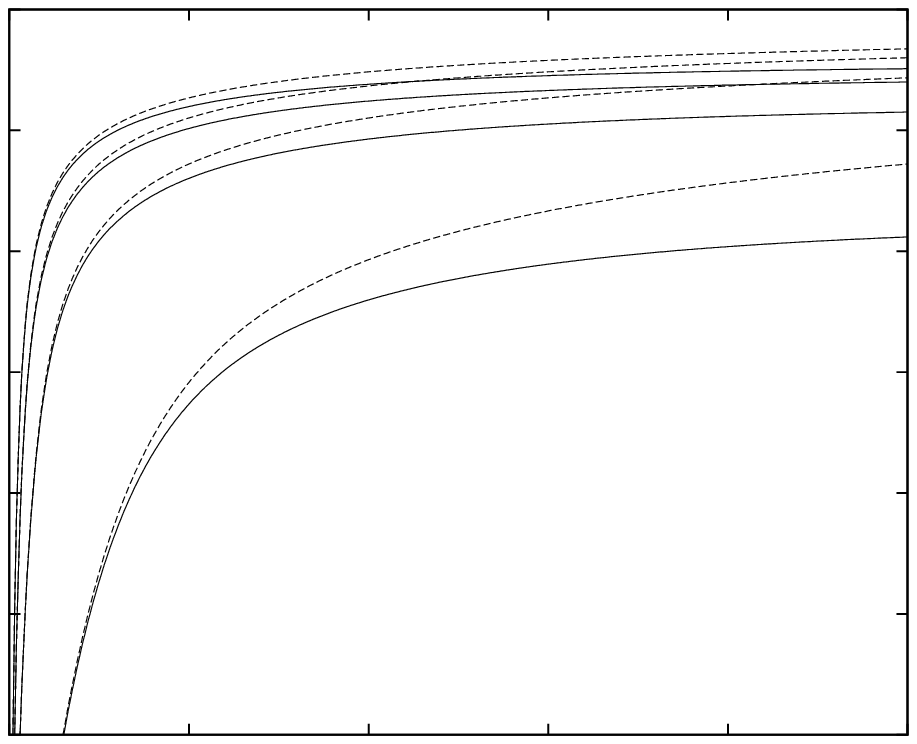}}%
    \gplfronttext
  \end{picture}%
\endgroup

%% file: witten_index.bbl
\begin{thebibliography}{99}
\bibitem{witten} E. Witten, 'Constraints on supersymmetry breaking', Nucl. Phys. B 202 (1982) 253.
\bibitem{claudson} M. Claudson, M.B. Halpern, 'Supersymmetric ground state wave functions', Nucl. Phys. B 250 (1985) 689.
\bibitem{hoppe} J. Hoppe, 'Quantum theory of a massless relativistic surface and a two dimensional bound state problem', PhD thesis at MIT (1982).
\bibitem{dewit} B. de Wit, J. Hoppe, H. Nicolai, 'On the quantum mechanics of supermembranes', Nucl. Phys. B 305 (1988) 545.
\bibitem{banks} T. Banks, W. Fischler, S. Shenker, L. Susskind, 'M-theory as a matrix model: a conjecture', Phys. Rev. D 55 (1997) 6189.
\bibitem{sethi+stern} S. Sethi, M. Stern, 'D-Brane bound states redux', Commun. Math. Phys 194 (1998) 675.
\bibitem{staudacher} M. Staudacher, 'Bulk Witten indices and the number of normalizable ground states
in supersymmetric quantum mechanics of orthogonal, symplectic and exceptional groups',  Phys. Lett. B 488 (2000) 194.
\bibitem{kanamori} I. Kanamori, 'A method for measuring the Witten index using lattice simulation', Nucl. Phys. B 841 (2010) 426.
\bibitem{doktorat_macka} M. Trzetrzelewski, 'Supersymmetric Yang-Mills quantum mechanics with arbitrary number of colors', PhD thesis at Jagiellonian University (2006).
\bibitem{wosiek1} J. Wosiek, 'Spectra of supersymmetric Yang-Mills quantum mechanics', Nucl. Phys. B 644 (2002) 85.
\bibitem{korcyl1} P. Korcyl, 'Recursive approach to supersymmetric quantum mechanics for arbitrary fermion occupation number', Acta Phys. Pol. B 41 (2010) 795.
\bibitem{korcyl4} P. Korcyl, 'Exact solutions to $D=2$, Supersymmetric Yang-Mills Quantum Mechanics with $SU(3)$ gauge group', Acta Phys. Pol. B Proc. Suppl. 2 (2009) 623.
\bibitem{korcyl5} P. Korcyl, 'Gauge invariant plane-wave solutions in supersymmetric Yang-Mills quantum mechanics', arXiv: 1008.2975.
\bibitem{korcyl6} P. Korcyl, 'Solutions of $D=2$ supersymmetric Yang-Mills quantum mechanics with $SU(N)$ gauge group', arXiv: 1101.0591.
\bibitem{shifman} M. A. Shifman, 'ITEP Lectures on Particle Physics and Field Theory', World Scientific vol. 1 (1999).
\bibitem{wosiek4} M. Campostrini, J. Wosiek, 'Exact Witten index in D=2 supersymmetric Yang-Mills quantum mechanics', Phys. Lett. B 550 (2002) 121.
\bibitem{dehesa} J.S. Dehesa, F. Dominguez Adame, E.R. Arriola, A. Zarzo, 'Hydrogen atom and orthogonal polynomials',
 IMACS Ann. Comput. Appl. Math. 9 (1991) 223,  Proc. III International Symposium on Orthogonal Polynomials and Their
Applications (Erice, Italy, June 1990).
\bibitem{dehesa2} R. Alvarez-Nodarse, J. S. Dehesa, 'Distributions of zeros of discrete and
continuous polynomials from their
recurrence relation', Appl. Math. and Comp. 128 (2002) 167.
\bibitem{knuth} R. L. Graham, D. E. Knuth, O. Patashnik, 'Concrete Mathematics', Addison Wesley (1990).
\bibitem{maciek2} M. Trzetrzelewski, J. Wosiek, 'Quantum systems in a cut Fock space', Acta Phys. Polon. B 35 (2004) 1615.
\bibitem{maciek_lie_groups} M. Trzetrzelewski, 'Supersymmetry and Lie groups', J. Math. Phys. 48 (2007) 083508.
%
%
%
%
%
%
%
%
%
\end{thebibliography}
